 \documentclass[sigconf]{acmart} 

\DeclareUnicodeCharacter{202F}{ }
\AtBeginDocument{%
  }

\acmConference[CHI '26 Workshop on Human-Agent Collaboration]{CHI '26 Workshop on Human-Agent Collaboration}{April 2026}{Barcelona, Spain}
\acmBooktitle{CHI '26 Workshop on Human-Agent Collaboration, April, 2026, Barcelona, Spain}

\usepackage{graphicx}
\usepackage{tikz}
\usepackage{color}
\usepackage{subcaption}
\usepackage{booktabs}
\usepackage{xcolor}
\usepackage{setspace}
\usepackage{tabularx}        
\usepackage{threeparttable}  
\usepackage{enumitem}
\usepackage{microtype}

\definecolor{mypurple}{RGB}{119, 69, 198}

\begin{document}

\title{Rethinking Health Agents: From Siloed AI to Collaborative Decision Mediators}





\author{Ray-Yuan Chung}
\email{raychung@uw.edu}
\affiliation{%
  \institution{University of Washington}
  \city{Seattle, WA}
    \country{USA}
}

\author{Xuhai “Orson” Xu}
\email{xx2489@cumc.columbia.edu}
\affiliation{%
  \institution{Columbia University}
  \city{New York, NY}
    \country{USA}
}

\author{Ari Pollack}
\email{ari.pollack@seattlechildrens.org}
\affiliation{%
  \institution{Seattle Children’s Hospital}
  \city{Seattle, WA}
    \country{USA}
}


\begin{abstract}

Large language model (LLM)–based health agents are increasingly used by health consumers and clinicians to interpret health information and guide health decisions. However, most AI systems in healthcare operate in siloed configurations, supporting individual users rather than the multi-stakeholder relationships central to healthcare. Such use can fragment understanding and exacerbate misalignment among patients, caregivers, and clinicians. We reframe AI not as a standalone assistant, but as a collaborator embedded within multi-party care interactions. Through a clinically validated fictional pediatric chronic kidney disease case study, we show that breakdowns in adherence stem from fragmented situational awareness and misaligned goals, and that siloed use of general-purpose AI tools does little to address these collaboration gaps. We propose a conceptual framework for designing AI collaborators that surface contextual information, reconcile mental models, and scaffold shared understanding while preserving human decision authority. 

\end{abstract}






\maketitle
\section{Introduction}
The rapid development of artificial intelligence (AI) has transformed how health consumers and clinicians access and interpret health information. Large language model (LLM)–based systems are increasingly used by consumers to seek medical advice and interpret symptoms \cite{Yun_Bickmore_2025, Paruchuri_etal_2025}, while generative AI tools such as OpenEvidence support clinicians in summarizing literature and assisting with diagnostic reasoning \cite{Hurt_2025}. As we move toward the era of personal health agents \cite{Heydari_2025}, AI is becoming embedded in everyday health decision-making. Despite advances in AI capabilities, current systems are largely designed for individual use and are not structured to support the multi-stakeholder interactions that characterize most clinical care. This parallel, siloed adoption raises important concerns: over-reliance on conversational agents may reduce critical evaluation of health information, particularly when responses are presented with high confidence, and clinicians have expressed concerns about accuracy, provenance, and accountability when such tools are used to search for or validate health-related information \cite{Bean_Payne_Parsons_etal_2026}.

Although prior work has explored AI systems that support multiple users \cite{Seo_Kim_Kim_Fan_Ackerman_Choi_Park_2025,Yang2024} in healthcare, these systems primarily facilitate information exchange rather than collaborative decision-making. For example, Talk2Care collects health information from older adults at home and transmits it to healthcare providers for clinical decision-making \cite{Yang2024}. Similarly, the ARCH chatbot gathers information from children and shares it with parents and providers \cite{Seo_Kim_Kim_Fan_Ackerman_Choi_Park_2025}. In these systems, AI primarily functions as an information conduit—collecting data from one party and passing it to another—without mediating perspectives, clarifying misunderstandings, or scaffolding shared reasoning across decision-makers. As a result, decision-making may still suffer from fragmented understanding, misaligned expectations, or invisible tensions among stakeholders.

\begin{figure*}[t]  
    \centering
    \Description{A side-by-side conceptual diagram of AI-agent collaboration. The left panel shows siloed use of AI chatbots by health consumers outside the clinic, highlighting issues such as over-reliance, distrust, and misaligned understanding between patients and clinicians. The right panel shows an AI collaborator positioned between health consumers and clinicians, with arrows indicating bidirectional communication and support for continuous care outside the clinic and improved collaboration during clinical visits.}
    \includegraphics[width=1\textwidth]{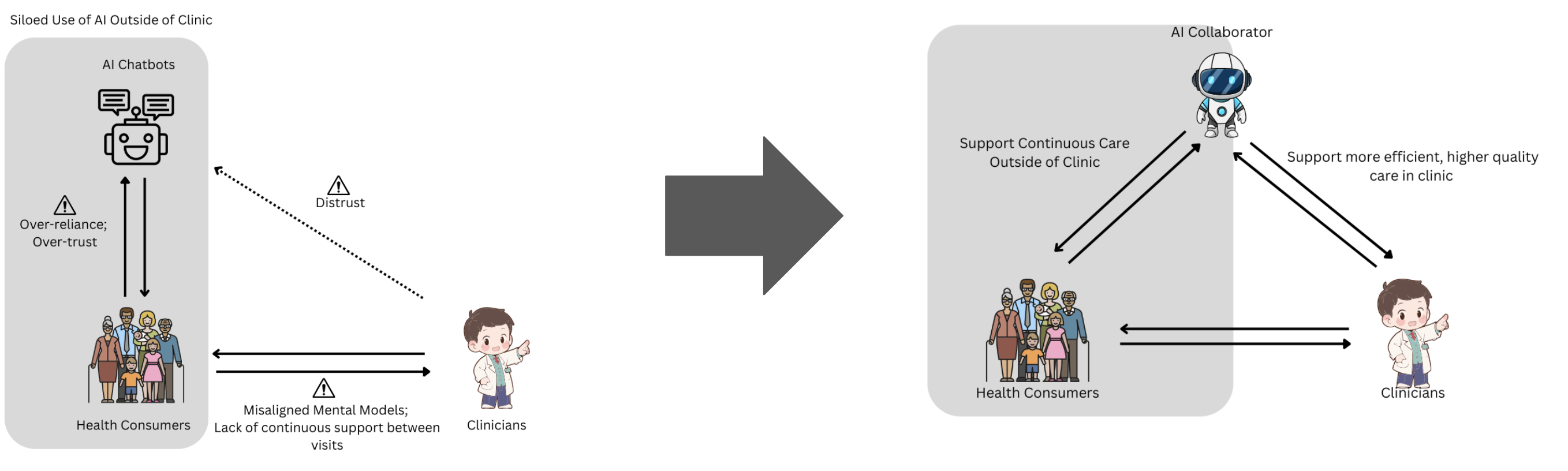}
    \caption{Proposed diagram comparing siloed AI use with an embedded AI collaborator in multi-stakeholder care.}
    \label{fig:HAI}
\end{figure*}

To address these limitations, we propose an AI collaborator framework designed to participate in and support multi-party healthcare decision-making, rather than serving a single end user (Figure~\ref{fig:HAI}). Building on prior multi-user systems, we move beyond information transmission toward supporting collaborative decision-making. In our conceptualization, the AI collaborator is embedded within triadic (or multi-party) care interactions and is designed to: 
\begin{itemize}
    \item Surface and contextualize relevant information among stakeholders, highlighting critical discrepancies, and information gaps.
    \item Identify and reconcile differing mental models, assumptions, or priorities among decision-makers.
    \item Scaffold shared understanding and coordinated reasoning processes to anticipate potential outcomes and implications of different decisions, while preserving human decision authority.
\end{itemize}

This shift in the role of AI raises critical design questions: What level of autonomy should such an AI possess? How should responsibility and accountability be distributed across stakeholders? And how can AI systems strengthen—rather than erode—professional judgment and patient autonomy?

To illustrate both the limitations of current healthcare interactions and the opportunities for AI-mediated collaboration, we present a fictional pediatric case study. This case highlights common barriers in today’s care models, including fragmented communication, mismatched expectations, incomplete contextual knowledge, and asymmetries in expertise. We then demonstrate how a deliberately designed AI collaborator could intervene—not as a decision-maker, but as a mediator that aligns stakeholders, clarifies trade-offs, and supports shared understanding. Through this conceptual and design exploration, we articulate principles for building AI collaborators as a carefully bounded participant in collaborative care.

\section{Design Scenario: A Pediatric Chronic Kidney Disease Case Study}

\subsection{Case Description}

Here, we present a fictional chronic kidney disease case study, validated by a pediatric nephrologist and a registered dietitian to ensure clinical relevance and ecological validity:

Alex, a 16-year-old high school student, was diagnosed with chronic kidney disease. During a clinic visit, the healthcare team explains that although Alex feels well, his blood pressure is elevated and dietary sodium intake must be reduced to protect kidney function. A strict low-sodium diet, daily medication, and regular monitoring are recommended. The teenager and his parent agree without raising concerns, trusting the clinician’s expertise and assuming that strict adherence is the only responsible course of action.

In the following weeks, dietary changes become the primary source of strain. At home, meals are carefully prepared to reduce sodium, but at school Alex struggles to avoid cafeteria food and snacks shared with friends. Reading nutrition labels feels tedious, and declining food in social settings feels isolating. In an effort to help, the parent turns to a general-purpose AI chatbot for low-sodium meal plans. While the system generates structured recommendations, the suggestions are generic and insufficiently tailored to the family’s cultural food preferences, budget constraints, and school schedule. As a result, many proposed meals feel impractical or unfamiliar. Small exceptions begin to accumulate—chips after practice, fast food with classmates—while these lapses are minimized at home. Tension grows as the parent reinforces the diet, and Alex becomes increasingly frustrated. By the next clinic visit, adherence has been inconsistent.

\subsection{Collaborative Decision-making Needs}

Healthcare decision-making has shifted from a clinician-driven model toward a collaborative approach that encourages greater patient participation \cite{Elwyn_Frosch_Thomson_Joseph-Williams_Lloyd_Kinnersley_Cording_Tomson_Dodd_Rollnick_etal_2012}. In pediatric chronic care, collaboration is critical because youth patients, caregivers, and clinicians must jointly navigate ongoing treatment decisions, yet such collaboration is often difficult due to their divergent goals, roles, and constraints \cite{Miller_2018}. In the chronic kidney disease case, the clinician prioritizes long-term risk reduction, the parent emphasizes compliance and safety, and the teenager values normalcy and autonomy. These differing priorities remain implicit during the visit and later surface as non-adherence to clinician recommendations and tension. Although the treatment plan is clinically appropriate, the breakdown still occurs due to misaligned goals and the absence of shared understanding of trade-offs and contextual constraints.

To conceptualize this challenge, we draw on Endsley’s team situational awareness (SA) framework \cite{Endsley_2016}. Originally developed in high-stakes domains such as aviation, team situational awareness (SA) is defined as the degree to which each team member possesses the situational awareness required for their responsibilities, aligned with a shared team goal and coordinated efforts to achieve that goal \cite{Endsley_2016}. This includes: (1) perception of relevant elements, (2) comprehension of their meaning, and (3) projection of their future status. Applied to pediatric chronic care, Level 1 includes access to relevant clinical indicators (e.g., blood pressure trends, sodium intake patterns), contextual factors (e.g., school routines, social pressures) and values of all decision-makers involved. Level 2 requires shared interpretation—understanding what those data mean medically and personally. Level 3 involves anticipating future outcomes, such as the long-term impact of choices or the sustainability of strict dietary rules. Prior work has shown that enhancing team situational awareness supports collaborative decision-making aimed at achieving shared goals in healthcare \cite{Pollack_Mishra_Apodaca_Khelifi_Haldar_Pratt_2020}. In our case, breakdowns occur because these three levels of awareness are fragmented across stakeholders rather than shared.

\subsection{AI Collaborator Design Consideration}

The team situational awareness framework provides a principled lens for designing an AI collaborator in this context. Rather than functioning as an autonomous decision-maker, an AI collaborator can be designed to strengthen shared awareness across stakeholders. Prior to appointments, the system could support low-burden meal logging, identify sodium intake patterns, and prompt reflection on situations where adherence feels most difficult (e.g., school lunches, team events). The AI could then synthesize these data into structured summaries that highlight both medical indicators and patient-reported challenges. During clinic visits, the system might surface trade-offs and simulate possible trajectories (e.g., projected blood pressure trends under different adherence scenarios), enabling projection and future-oriented discussion.

This design contrasts fundamentally from the siloed use of gen\-er\-al-pur\-pose AI tools. When patients or caregivers independently consult AI chatbots, the resulting advice is often decontextualized, disconnected from clinical records, and invisible to clinicians. Such siloed interactions can introduce inconsistencies, reinforce misunderstandings, or create parallel decision pathways that undermine collaborative decision-making. In contrast, an AI collaborator embedded within the care relationship generates shared artifacts visible to all stakeholders, mediates perspectives, and aligns contextual realities with clinical goals. The clinician retains decision authority, and the teenager retains agency; the AI supports sustainable collaboration without replacing human judgment.

\section{Future Directions}

We invite the HCI community to further explore the following research directions for future AI collaborators in healthcare.

(1) \textbf{In different health-related contexts and tasks, what levels of agent autonomy are preferred by decision-makers?} Across domains like software engineering, and autonomous vehicles, we are observing a broader shift from AI as passive tools toward AI as active collaborators—and in some contexts, even toward configurations where humans become observers of automated systems \cite{Feng2025}. Healthcare, however, presents distinct ethical and practical constraints. Licensed clinicians and patients are expected to retain final decision-making authority, especially patients must retain agency over their care. In addition,  patients of different ages vary in their desire and readiness to take responsibility for their care \cite{Dunbar_Bascom_Pratt_Snyder_Smith_Pollack_2022}. Therefore, designing AI collaborators therefore requires careful calibration of autonomy—balancing guidance, delegation, and oversight—based on developmental stage, task complexity, and stakeholder roles.

(2) \textbf{When AI col\-lab\-o\-ra\-tors are introduced into multi-stake\-hold\-er care teams, do they mitigate or exacerbate tensions in human–human relationships?}  
For example, how should systems respond when AI-generated suggestions conflict with clinician recommendations or caregiver preferences? These questions apply to any context involving multiple parties, particularly caregivers who play active roles in decision making, including pediatrics, mental health \cite{Schuster2021, Tuijt2021}, and long term care \cite{Niedling2025, Koster2021}. Prior work shows that mismatched expectations regarding caregiver participation versus patient autonomy often create friction in collaborative decision-making \cite{Boland_Graham_Légaré_Lewis_Jull_Shephard_Lawson_Davis_Yameogo_Stacey_2019,Hong_Wilcox_Machado_Olson_Simoneaux_2016}. AI collaborators must therefore be designed with explicit mechanisms for transparency, role clarity, and conflict mediation, while accounting for evolving patient–caregiver dynamics across developmental stages \cite{Toscos_Connelly_Rogers_2012}.

\section{Conclusion}
In this work, we reframed AI in healthcare not as a standalone assistant, but as a collaborator embedded within multi-stakeholder care relationships. Through a fictional pediatric chronic kidney disease case study grounded in clinical validation, we illustrated how breakdowns in adherence stem from fragmented situational awareness and misaligned goals across patients, caregivers, and clinicians. Drawing on team situational awareness theory, we outlined how AI collaborators can strengthen shared perception, comprehension, and projection without displacing human authority or autonomy. By positioning AI as a mediator that surfaces contextual information, reconciles mental models, and scaffolds longitudinal decision-making, this work offers a conceptual foundation for designing ethically grounded collaborative AI systems in healthcare.



\bibliographystyle{ACM-Reference-Format}
\bibliography{CHI26}

@article{Hurt_2025,
  author    = {Hurt, R. T. and Stephenson, C. R. and Gilman, E. A. and Aakre, C. A. and Croghan, I. T. and Mundi, M. S. and Ghosh, K. and Edakkanambeth Varayil, J.},
  title     = {The Use of an Artificial Intelligence Platform OpenEvidence to Augment Clinical Decision-Making for Primary Care Physicians},
  journal   = {Journal of Primary Care \& Community Health},
  year      = {2025},
  volume    = {16},
  pages     = {21501319251332215},
  doi       = {10.1177/21501319251332215},
  pmid      = {40238861},
  pmcid     = {PMC12033599},
  note      = {Epub 2025 Apr 16}
}

@article{Bean_Payne_Parsons_etal_2026,
  author    = {Bean, Andrew M. and Payne, Rebecca E. and Parsons, Guy and Kirk, Hannah R. and Ciro, Juan and Mosquera-G{\'o}mez, Rafael and Hincapi{\'e}, Sara and Ekanayaka, Aruna S. and Tarassenko, Lionel and Rocher, Luc and Mahdi, Adam},
  title     = {Reliability of {LLMs} as medical assistants for the general public: a randomized preregistered study},
  journal   = {Nature Medicine},
  year      = {2026},
  volume    = {32},
  number    = {3},
  pages     = {654--662},
  doi       = {10.1038/s41591-025-04074-y},
  url       = {https://www.nature.com/articles/s41591-025-04074-y}
}

@misc{Heydari_2025,
      title={The Anatomy of a Personal Health Agent}, 
      author={A. Ali Heydari and Ken Gu and Vidya Srinivas and Hong Yu and Zhihan Zhang and Yuwei Zhang and Akshay Paruchuri and Qian He and Hamid Palangi and Nova Hammerquist and Ahmed A. Metwally and Brent Winslow and Yubin Kim and Kumar Ayush and Yuzhe Yang and Girish Narayanswamy and Maxwell A. Xu and Jake Garrison and Amy Armento Lee and Jenny Vafeiadou and Ben Graef and Isaac R. Galatzer-Levy and Erik Schenck and Andrew Barakat and Javier Perez and Jacqueline Shreibati and John Hernandez and Anthony Z. Faranesh and Javier L. Prieto and Connor Heneghan and Yun Liu and Jiening Zhan and Mark Malhotra and Shwetak Patel and Tim Althoff and Xin Liu and Daniel McDuff and Xuhai "Orson" Xu},
      year={2025},
      eprint={2508.20148},
      archivePrefix={arXiv},
      primaryClass={cs.AI},
      url={https://arxiv.org/abs/2508.20148}, 
}

@inproceedings{Paruchuri_etal_2025,
    title = "``What{'}s Up, Doc?'': Analyzing How Users Seek Health Information in Large-Scale Conversational {AI} Datasets",
    author = "Paruchuri, Akshay  and
      Aziz, Maryam  and
      Vartak, Rohit  and
      Ali, Ayman  and
      Uchehara, Best  and
      Liu, Xin  and
      Chatterjee, Ishan  and
      Agrawal, Monica",
    editor = "Christodoulopoulos, Christos  and
      Chakraborty, Tanmoy  and
      Rose, Carolyn  and
      Peng, Violet",
    booktitle = "Findings of the Association for Computational Linguistics: EMNLP 2025",
    month = nov,
    year = "2025",
    address = "Suzhou, China",
    publisher = "Association for Computational Linguistics",
    url = "https://aclanthology.org/2025.findings-emnlp.125/",
    doi = "10.18653/v1/2025.findings-emnlp.125",
    pages = "2312--2336",
    ISBN = "979-8-89176-335-7"
}

@article{Yun_Bickmore_2025,
  author       = {Yun, H. and Bickmore, T.},
  title        = {Online Health Information–Seeking in the Era of Large Language Models: Cross-Sectional Web-Based Survey Study},
  journal      = {Journal of Medical Internet Research},
  year         = {2025},
  volume       = {27},
  pages        = {e68560},
  doi          = {10.2196/68560},
  url          = {https://www.jmir.org/2025/1/e68560}
}

@misc{Feng2025,
      title={Levels of Autonomy for AI Agents}, 
      author={K. J. Kevin Feng and David W. McDonald and Amy X. Zhang},
      year={2025},
      eprint={2506.12469},
      archivePrefix={arXiv},
      primaryClass={cs.HC},
      url={https://arxiv.org/abs/2506.12469}, 
}

@article{Yang2024,
author = {Yang, Ziqi and Xu, Xuhai and Yao, Bingsheng and Rogers, Ethan and Zhang, Shao and Intille, Stephen and Shara, Nawar and Gao, Guodong Gordon and Wang, Dakuo},
title = {Talk2Care: An LLM-based Voice Assistant for Communication between Healthcare Providers and Older Adults},
year = {2024},
issue_date = {June 2024},
publisher = {Association for Computing Machinery},
address = {New York, NY, USA},
volume = {8},
number = {2},
url = {https://doi.org/10.1145/3659625},
doi = {10.1145/3659625},
journal = {Proc. ACM Interact. Mob. Wearable Ubiquitous Technol.},
month = may,
articleno = {73},
numpages = {35}
}

@article{Koster2021,
  author  = {Koster, Luzan and Verbeek, Hilde and de Boer, Bernadette and Hamers, Jan P. H.},
  title   = {It takes three to tango: An ethnography of triadic involvement of residents, families and nurses in long-term dementia care},
  journal = {Health Expectations},
  year    = {2021},
  volume  = {24},
  number  = {4},
  pages   = {1305--1315},
  doi     = {10.1111/hex.13224}
}

@article{Niedling2025,
  author  = {Niedling, Katharina and Richter, Stefanie and H\"{a}mel, Kerstin},
  title   = {Triadic relationships in home care nursing: an integrative review of the views and experiences of older couples and nurses},
  journal = {BMC Nursing},
  year    = {2025},
  volume  = {24},
  number  = {1},
  pages   = {671},
  doi     = {10.1186/s12912-025-03378-1},
  note    = {Published 2025 Jun 27}
}

@article{Tuijt2021,
  author  = {Tuijt, R. and Rees, J. and Frost, R. and Wilcock, J. and Manthorpe, J. and Rait, G.},
  title   = {Exploring how triads of people living with dementia, carers and health care professionals function in dementia health care: A systematic qualitative review and thematic synthesis},
  journal = {Dementia (London)},
  year    = {2021},
  volume  = {20},
  number  = {3},
  pages   = {1080--1104},
  doi     = {10.1177/1471301220915068}
}

@article{Schuster2021,
  author = {Schuster, F. and Holzhüter, F. and Heres, S. and Hamann, J.},
  title = {'Triadic' shared decision making in mental health: Experiences and expectations of service users, caregivers and clinicians in Germany},
  journal = {Health Expectations},
  year = {2021},
  volume = {24},
  number = {2},
  pages = {507--515},
  doi = {10.1111/hex.13192}
}

@article{Elwyn_Frosch_Thomson_Joseph-Williams_Lloyd_Kinnersley_Cording_Tomson_Dodd_Rollnick_etal_2012,
  title   = {Shared Decision Making: A Model for Clinical Practice},
  author  = {Elwyn, Glyn and Frosch, Dominick and Thomson, Richard and Joseph-Williams, Natalie and Lloyd, Amy and Kinnersley, Paul and Cording, Emma and Tomson, Dave and Dodd, Carole and Rollnick, Stephen and Edwards, Adrian and Barry, Michael},
  journal = {Journal of General Internal Medicine},
  volume  = {27},
  number  = {10},
  pages   = {1361--1367},
  year    = {2012},
  month   = {oct},
  doi     = {10.1007/s11606-012-2077-6},
  issn    = {0884-8734, 1525-1497},
  language= {en}
}

@article{Boland_Graham_Légaré_Lewis_Jull_Shephard_Lawson_Davis_Yameogo_Stacey_2019, title={Barriers and facilitators of pediatric shared decision-making: a systematic review}, volume={14}, ISSN={1748-5908}, DOI={10.1186/s13012-018-0851-5}, abstractNote={Background: Shared decision-making (SDM) is rarely implemented in pediatric practice. Pediatric health decisionmaking differs from that of adult practice. Yet, little is known about the factors that influence the implementation of pediatric shared decision-making (SDM). We synthesized pediatric SDM barriers and facilitators from the perspectives of healthcare providers (HCP), parents, children, and observers (i.e., persons who evaluated the SDM process, but were not directly involved). Methods: We conducted a systematic review guided by the Ottawa Model of Research Use (OMRU). We searched MEDLINE, EMBASE, Cochrane Library, CINAHL, PubMed, and PsycINFO (inception to March 2017) and included studies that reported clinical pediatric SDM barriers and/or facilitators from the perspective of HCPs, parents, children, and/or observers. We considered all or no comparison groups and included all study designs reporting original data. Content analysis was used to synthesize barriers and facilitators and categorized them according to the OMRU levels (i.e., decision, innovation, adopters, relational, and environment) and participant types (i.e., HCP, parents, children, and observers). We used the Mixed Methods Appraisal Tool to appraise study quality. Results: Of 20,008 identified citations, 79 were included. At each OMRU level, the most frequent barriers were features of the options (decision), poor quality information (innovation), parent/child emotional state (adopter), power relations (relational), and insufficient time (environment). The most frequent facilitators were low stake decisions (decision), good quality information (innovation), agreement with SDM (adopter), trust and respect (relational), and SDM tools/resources (environment). Across participant types, the most frequent barriers were insufficient time (HCPs), features of the options (parents), power imbalances (children), and HCP skill for SDM (observers). The most frequent facilitators were good quality information (HCP) and agreement with SDM (parents and children). There was no consistent facilitator category for observers. Overall, study quality was moderate with quantitative studies having the highest ratings and mixedmethod studies having the lowest ratings. Conclusions: Numerous diverse and interrelated factors influence SDM use in pediatric clinical practice. Our findings can be used to identify potential pediatric SDM barriers and facilitators, guide context-specific barrier and facilitator assessments, and inform interventions for implementing SDM in pediatric practice. Trial Registration: PROSPERO CRD42015020527}, number={1}, journal={Implementation Science}, author={Boland, Laura and Graham, Ian D. and Légaré, France and Lewis, Krystina and Jull, Janet and Shephard, Allyson and Lawson, Margaret L. and Davis, Alexandra and Yameogo, Audrey and Stacey, Dawn}, year={2019}, month=dec, pages={7}, language={en} }

@article{Dunbar_Bascom_Pratt_Snyder_Smith_Pollack_2022, title={My Kidney Identity: Contextualizing pediatric patients and their families kidney transplant journeys}, volume={26}, ISSN={1399-3046}, DOI={10.1111/petr.14343}, abstractNote={BACKGROUND: Even though having a kidney transplant is the treatment of choice for children with kidney failure, it can cause anxiety for patients and their families resulting in decreased psychosocial functioning, adherence, and self-management. We set out to identify the information needs required to help pediatric patients and their families contextualize their posttransplant experiences as they recalibrate their understanding of normalcy throughout their transplant journey. METHODS: Participants submitted photographs related to feeling: (1) worried, (2) confident, (3) similar to peers without kidney disease, and (4) different from these peers. The photographs served as a foundation for an in-depth interview. RESULTS: Nineteen individuals (10 pediatric transplant recipients and 9 caregivers) were interviewed at a mean of 8 years posttransplant. We identified five specific themes and tensions our participants associated with recalibrating their version of “normal” throughout the transplant journey: (1) exchanging information (information consumers vs. information contributors, (2) transitional management (family management vs. self-management), (3) building confidence (worry vs. confidence), (4) telling one’s story (hiding vs. self-expression), and (5) normalizing kidney transplantation (feeling different vs. feeling similar). These five themes/tensions form one’s Kidney Identity, shift from negative to positive throughout the transplant journey, illustrating a more abstract and complex account of kidney transplantation over time. CONCLUSIONS: Having a patient view their Kidney Identity over time may support self-reflection of one’s progress posttransplant and potentially help clinicians, patients, and their caregivers identify barriers and areas where they may need more support to ensure their successful engagement in their care.}, number={7}, journal={Pediatric Transplantation}, author={Dunbar, Julia C. and Bascom, Emily and Pratt, Wanda and Snyder, Jaime and Smith, Jodi M. and Pollack, Ari H.}, year={2022}, month=nov, pages={e14343}, language={eng} }

@book{Endsley_2016, address={Boca Raton}, edition={2}, title={Designing for Situation Awareness: An Approach to User-Centered Design, Second Edition}, ISBN={978-0-429-14673-2}, DOI={10.1201/b11371}, abstractNote={Liberally illustrated with actual design examples, this book demonstrates how people acquire and interpret information and examines the factors that undermine this process. The second edition expands and updates the examples throughout to include a wider range of domains and increases the coverage of SA design principles and guidelines to include new areas of development.}, publisher={CRC Press}, author={Endsley, Mica R.}, year={2016}, month=apr }

@article{Hong_Wilcox_Machado_Olson_Simoneaux_2016, title={Care Partnerships: Toward Technology to Support Teens’ Participation in Their Health Care}, volume={2016}, DOI={10.1145/2858036.2858508}, abstractNote={Adolescents with complex chronic illnesses, such as cancer and blood disorders, must partner with family and clinical caregivers to navigate risky procedures with life-altering implications, burdensome symptoms and lifelong treatments. Yet, there has been little investigation into how technology can support these partnerships. We conducted 38 in-depth interviews (15 with teenage adolescents with chronic forms of cancer and blood disorders, 15 with their parents, and eight with clinical caregivers) along with nine non-participant observations of clinical consultations to better understand common challenges and needs that could be supported through design. Participants faced challenges primarily concerning: 1) teens’ limited participation in their care, 2) communicating emotionally-sensitive information, and 3) managing physical and emotional responses. We draw on these findings to propose design goals for sociotechnical systems to support teens in partnering in their care, highlighting the need for design to support gradually evolving partnerships in care.}, journal={Proceedings of the SIGCHI conference on human factors in computing systems . CHI Conference}, author={Hong, Matthew K. and Wilcox, Lauren and Machado, Daniel and Olson, Thomas A. and Simoneaux, Stephen F.}, year={2016}, month=may, pages={5337–5349} }

@inproceedings{Seo_Kim_Kim_Fan_Ackerman_Choi_Park_2025, address={Yokohama Japan}, title={Enhancing Pediatric Communication: The Role of an AI-Driven Chatbot in Facilitating Child-Parent-Provider Interaction}, ISBN={979-8-4007-1394-1}, url={https://dl.acm.org/doi/10.1145/3706598.3713134}, DOI={10.1145/3706598.3713134}, abstractNote={Communication with child patients is challenging due to their developing ability to express emotions and symptoms. Additionally, healthcare providers often have limited time to ofer resources to parents. By leveraging AI to facilitate free-form conversations, our study aims to design an AI-driven chatbot to bridge these gaps in child-parent-provider communication. We conducted two studies: 1) design sessions with 12 children with cancer and their parents, which informed the development of our chatbot, ARCH, and 2) an interview study with 15 pediatric care experts to identify potential challenges and refne ARCH’s role in pediatric communication. Our fndings highlight three key roles for ARCH: providing an expressive outlet for children, ofering reassurance to parents, and serving as an assessment tool for providers. We conclude by discussing design considerations for AI-driven chatbots in pediatric communication, such as creating communication spaces, balancing the expectations of children and parents, and addressing potential cultural diferences.}, booktitle={Proceedings of the 2025 CHI Conference on Human Factors in Computing Systems}, publisher={ACM}, author={Seo, Woosuk and Kim, Young-Ho and Kim, Ji Eun and Fan, Megan Tao and Ackerman, Mark S. and Choi, Sung Won and Park, Sun Young}, year={2025}, month=apr, pages={1–16}, language={en} }

\end{document}